\documentclass[10pt,aps,prb,reprint,longbibliography]{revtex4-1}   % style for Physical Review B and AJP are similar

\usepackage{amsmath}    % need for subequations
\usepackage{graphicx}   % for figures
\include{epsf}

\newcommand{\imu}{\text{i}} % to get imaginary unit upright
\newcommand{\diff}{\text{d}} % to get differential operator upright
\newcommand{\vect}[1]{\boldsymbol{#1}} % to get vectors in boldface
\newcommand{\dd}[2]{\frac{\text{d}^2 #1}{\text{d} #2^2}} % for double derivatives
\newcommand{\ket}[1]{\left| #1 \right>} % for Dirac bras
\newcommand{\bra}[1]{\left< #1 \right|} % for Dirac kets

\newcommand{\commentOut}[1]{}
\newcommand{\figwidth}{0.9\columnwidth}

\usepackage{times}
\renewcommand{\d}[2]{\frac{\text{d} #1}{\text{d} #2}} % for derivatives

\begin{document}

\title{The Classical Bloch Equations}
\author{Martin Frimmer and Lukas Novotny}
 \affiliation{ETH Z{\"u}rich, Photonics Laboratory, 8093 Z{\"u}rich, Switzerland.}
 \email{www.nano-optics.org}   %optional
\date{\today}

\begin{abstract}
Coherent control of a quantum mechanical two-level system is at the heart of magnetic resonance imaging, quantum information processing, and quantum optics. Among the most prominent phenomena in quantum coherent control are Rabi oscillations, Ramsey fringes and Hahn echoes. We demonstrate that these phenomena can be derived classically by use of a simple coupled harmonic oscillator model. The classical problem can be cast in a form that is formally equivalent to the quantum mechanical Bloch equations
with the exception that the longitudinal and the transverse relaxation times ($T_1$ and $T_2$) are equal.
% decay rate of the inversion ($1/T_1$) and the dephasing rate ($1/T_2$) turn out to be equal.
The classical analysis is intuitive and well suited for familiarizing students with the basic concepts of quantum coherent control, while at the same time highlighting the fundamental differences between classical and quantum theories. \\
\end{abstract}

\maketitle

\section{Introduction}
The harmonic oscillator is arguably the most fundamental building block at the core of our understanding of both classical and quantum physics. Interestingly, a host of phenomena originally encountered in quantum mechanics and initially thought to be of purely quantum-mechanical nature have been successfully modelled using coupled classical harmonic oscillators. Amongst these phenomena are electromagnetically induced transparency~\cite{Alzar2002}, rapid adiabatic passage~\cite{Shore2009,Maris1988} and Landau-Zener tunneling.\cite{Novotny2010}
A particularly rich subset of experiments is enabled by the coherent manipulation of a quantum mechanical two-level system, providing access to fascinating effects including Rabi oscillations, Ramsey fringes and Hahn echoes.\cite{Allen1987}
Remarkably, equipped with the models and ideas gained from studying quantum mechanical systems, researchers have returned to construct classical analogues of two-level systems.\cite{Dragoman2004,Tobar1991}
Coherent control of such a classical two-level system has been beautifully demonstrated for  an ``optical atom'' consisting of two coupled modes of a cavity.\cite{Spreeuw1990,Bouwmeester1995} Recently, coherent control of classical two-level systems has been achieved with coupled micromechanical oscillators.\cite{Okamoto2013,Faust2013}
With the analogy between a two-level system and a coupled pair of classical harmonic oscillators well established, it is surprising that this analogy has not been used to familiarize students with the concepts of coherent control and to provide an accessible analogue to a variety of quantum optical phenomena. Furthermore, exploring the limits of any analogue typically illustrates very strikingly the genuine features of a physical theory which are not present in the theory in which the analogy is phrased.\cite{Frimmer2012}\\[-1ex]

In this paper we consider a pair of two parametrically driven coupled harmonic mechanical oscillators. From the Newtonian equations of motion we derive a set of equations of motion for the eigenmode amplitudes that are formally equivalent to the time-dependent Schr\"{o}dinger equation of a two-level atom. We then derive a set of coupled differential equations which are formally identical with the quantum Bloch equations, with the exception that the longitudinal and transverse relaxation times are equal. We illustrate the concept of the Bloch sphere  and provide an intuitive understanding of coherent control experiments by discussing Rabi oscillations, Ramsey fringes and Hahn echoes. Finally, we point out the distinct differences between our mechanical analogue and a true quantum mechanical two-level system.
Our approach offers students an intuitive entry into the field and prepares them with the basic concepts of quantum coherent control.\\[-1ex]

\section{The mechanical atom}
\subsection{Equations of motion}
\label{sec:coupledosc}

\begin{figure}[!b]
\includegraphics[width=\figwidth]{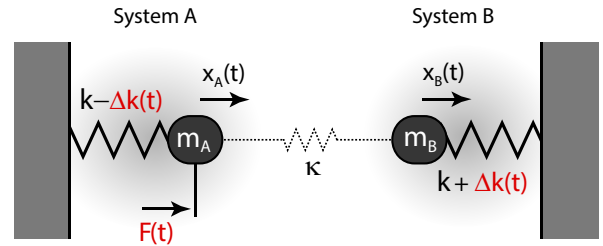}
%\epsfysize=10em
%\centerline{\epsfxsize=4cm \epsfbox{./Fig_coupledoscillators.eps}}
\caption{Coupled mechanical oscillators with masses $m_A, m_B$ and spring constants $k_A=k-\Delta k, k_B=k+\Delta k$ with a detuning $\Delta k$ that can be time dependent. The oscillators are coupled with a spring of spring constant $\kappa$. Oscillator A can be driven by an external force $F(t)$.
\label{fig:coupledoscillators}}
\end{figure}

Throughout this paper we consider two coupled oscillators, as illustrated in Fig.~\ref{fig:coupledoscillators}, with masses  $m_A$ and $m_B$, spring constants $k_A=k-\Delta k(t)$ and $k_B=k+\Delta k(t)$ with a small detuning $\Delta k(t)$ that can be time dependent, and coupled by a spring with spring constant $\kappa$, which is weak compared to $k$. Both oscillators are weakly damped at a rate $\gamma$. Oscillator $A$ can be externally driven by a force $F(t)$. Because of the coupling $\kappa$, the dynamics of oscillator $A$ couples over to oscillator $B$.
Such two coupled harmonic oscillators are a generic model system applicable to diverse fields of physics. For example, in molecular physics, oscillators $A$ and $B$ correspond to a pair of atoms. Similarly, in cavity quantum electrodynamics, $A$ is a two-level atom and $B$ is a cavity field. In cavity optomechanics, oscillator $A$ would represent a mechanical oscillator, such as a membrane or cantilever, and $B$ an optical resonator. For the following, we assume that the masses of the oscillators are equal ($m_A = m_B = m$). Then, in terms of the coordinates $x_A$ and $x_B$ of the two oscillators, the equations of motion are
\begin{equation}
\label{eq:coupledosc}
\begin{aligned}
\ddot{x}_A \,+\, \gamma\:\! \dot{x}_A \,+\, \left[\frac{k+\kappa}{m} - \frac{\Delta k(t)}{m}\right] x_A \,-\, \frac{\kappa}{m}\:\! x_B &\;=\; \frac{1}{m} F(t) \\[1ex]
\ddot{x}_B \,+\, \gamma\:\! \dot{x}_B \,+\, \left[\frac{k+\kappa}{m} + \frac{\Delta k(t)}{m}\right] x_B \,-\, \frac{\kappa}{m}\:\! x_A &\;=\; 0 .
\end{aligned}
\end{equation}
%If oscillator ${\rm B}$ were held at rest ($x_B=0$) and no external forces were applied ($\delta k=0, F=0$), oscillator  ${\rm A}$ would oscillate with a frequency $[\Omega_0^2+\tilde\Omega^2]^{1/2}$. {\it Vice versa}, oscillator  ${\rm B}$ would oscillate with a frequency $[\Omega_0^2-\tilde\Omega^2]^{1/2}$ if oscillator ${\rm A}$ were held at rest. \\

For ease of notation, we introduce the carrier frequency $\Omega_0$, the detuning frequency $\Omega_d$ and the coupling frequency $\Omega_c$ as
\begin{equation}
\label{eq:frequencies}
\begin{aligned}
\Omega_0^2&= [k+\kappa] / m\\
\Omega_d^2&= \Delta k / m\\
%\delta\Omega(t)^2&=& \delta k(t) \:\!/\:\! 2m  \nonumber\\
\Omega_c^2&= \kappa/m
\end{aligned}
 \end{equation}
and represent the coupled differential equations in Eq.~\eqref{eq:coupledosc} in matrix form as
\begin{equation}
\label{eq:coupledoscMatrixForm}
\left[\dd{}{t} + \gamma \d{}{t} + \Omega_0^2\right]\!
\begin{bmatrix} x_A \\ x_B \end{bmatrix} + \begin{bmatrix} -\Omega_d^2 & - \Omega_c^2\\ -\Omega_c^2 & \;\,\Omega_d^2\end{bmatrix}\begin{bmatrix} x_A \\ x_B \end{bmatrix}
\;=\; \begin{bmatrix} f(t) \\ 0 \end{bmatrix},
\end{equation}
where $f(t) = F(t)/m$. This system of equations describes the full dynamics of the coupled oscillator problem.\\[-1ex]

\subsection{Eigenmodes for constant detuning}
We first consider the case of constant detuning $\Delta k=\mathrm{const.}$ and solve for the eigenmodes of the system and their respective eigenfrequencies. To this end, we diagonalize the matrix in Eq.~\eqref{eq:coupledoscMatrixForm}. The eigenmodes  $x_{e1}$ and $x_{e2}$ of the system can be derived from the coordinates of the two oscillators $x_A, x_B$ as
\begin{equation}
\label{eq:transformEigenmodes}
\begin{bmatrix} x_A\\x_B\end{bmatrix} = U^{-1}\begin{bmatrix} x_{e1} \\ x_{e2} \end{bmatrix},
\end{equation}
where $U$ is a  transformation matrix whose rows are  eigenvectors of the matrix in Eq.~\eqref{eq:coupledoscMatrixForm}.  We find
\begin{equation}
U = \begin{bmatrix} U_{11} & U_{12}\\ U_{21} & U_{22} \end{bmatrix} = \begin{bmatrix} 1 & \;\;-(\Omega_d / \Omega_c)^2 + \sqrt{1+ (\Omega_d / \Omega_c)^4} \\
1 & \;\;-(\Omega_d / \Omega_c)^2 - \sqrt{1+ (\Omega_d / \Omega_c)^4} \end{bmatrix} \; ,
\end{equation}
and the eigenfrequencies turn out to be
\begin{equation}
\Omega_{\pm} =  \left[\Omega_0^2 \mp \sqrt{\Omega_d^4+\Omega_c^4}\,\right]^{1/2} \; .
\end{equation}
Thus, after transformation, Eq.~\eqref{eq:coupledoscMatrixForm} yields two independent differential equations for the normal mode coordinates $x_{e1}$ and $x_{e2}$
\begin{equation}
\begin{aligned}
\left[\dd{}{t} + \gamma \d{}{t} + \Omega_+^2\right] x_{e1} &=& U_{11} \,f (t) \\[1ex]
\left[\dd{}{t} + \gamma \d{}{t} + \Omega_-^2\right] x_{e2} &=& U_{21} \,f (t).
\end{aligned}
\label{eq:decoupledHarmOsc}
\end{equation}
\begin{figure*}[htb]
\includegraphics[width=1.4\columnwidth]{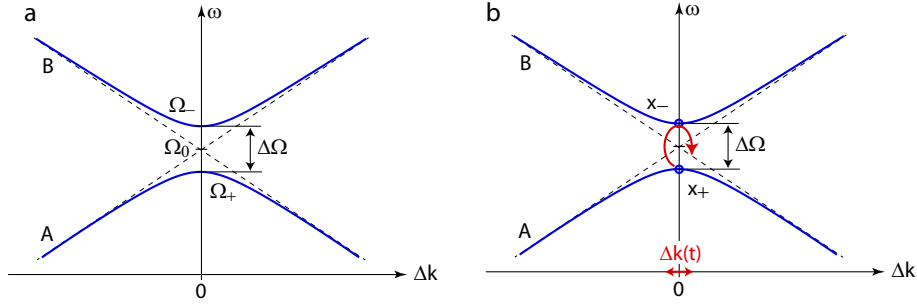}
%\epsfysize=15em
%\centerline{\epsfbox{./Fig_splitting.eps}}
\caption{(a) Eigenfrequencies $\Omega_+$ and $\Omega_-$ of the coupled oscillators as a function of the detuning $\Delta k$. The dashed lines show the eigenfrequencies in absence of coupling. The frequency splitting $\Delta\Omega$ at resonance ($\Delta k=0$) is proportional to the coupling strength $\kappa$. (b) The energy of the system can be swapped  between the eigenmodes by harmonically modulating the spring detuning $\Delta k$.
\label{fig:splitting}}
\end{figure*}
In Fig.~\ref{fig:splitting}a we plot the eigenfrequencies $\Omega_{\pm}$ as a function of the detuning $\Delta k$.  In the absence of coupling  the two oscillators are independent and their eigenfrequencies follow the straight lines that intersect at $\Delta k = 0$ in Fig.~\ref{fig:splitting}a. However, in presence of finite coupling, the two curves no longer intersect. Instead, there is a characteristic  anti-crossing of the eigenfrequencies. The frequency splitting at resonance ($\Delta k = 0$) is
\begin{equation}
\Delta\Omega\;=\; \Omega_--\Omega_+ \approx \frac{\Omega_c^2}{\Omega_0},
\end{equation}
where we made use of $\Omega_c\ll\Omega_0$. Thus, the splitting is proportional to the coupling strength $\kappa$. If the separation of the frequency branches $\Delta\Omega$ can be discriminated against their width, which scales with the damping constant $\gamma$, one considers the system to be in the so-called {\em strong coupling} regime.\cite{Novotny2010} Turning to the eigenmodes of the system, we find that on resonance the transformation matrix reads
\begin{equation}
U (\Delta k = 0) \;=\; \begin{bmatrix} 1 & 1\\ 1 & -1 \end{bmatrix}  \label{eq:TrafoUres} \; ,
\end{equation}
and the eigenmodes become
\begin{equation}
\begin{aligned}
\label{eq:Res_eigenmodes}
x_+ &=& \left. x_{e1}\right|_{\Delta k=0} &\,=\, x_A + x_B\\
x_- &=& \left. x_{e2}\right|_{\Delta k=0} &\,=\, x_A - x_B .
\end{aligned}
\end{equation}
Thus, on resonance, the eigenmodes of the system are symmetric and antisymmetric superpositions of the two individual oscillators. For $x_+$ the two masses swing in phase and for $x_-$ out of phase, that is, against each other. The eigenfrequency of the symmetric mode is $\Omega_+=\sqrt{k\,/\,m}$, which is the frequency in absence of coupling. This result is obvious since the coupling spring plays no role for the symmetric mode. The eigenfrequency of the antisymmetric mode is $\Omega_-=\sqrt{(k+2\kappa)\,/\,m}$. It is higher than $\Omega_+$ because each oscillator feels the coupling spring.\\[-1ex]

We have thus far considered a static detuning $\Delta k$. Intriguing effects happen when $\Delta k$ becomes  time dependent. For example, if oscillator $A$ is excited and the detuning $\Delta k$ is  swept through the anti-crossing region, then, depending on how fast the parameter $\Delta k$ is varied, one can transfer the  energy to oscillator $B$ or keep it on oscillator $A$. The former is referred to as an  adiabatic transition and the latter as a diabatic transition. In a diabatic transition the system jumps from one branch in Fig.~\ref{fig:splitting}a to the other, a process referred to as a Landau-Zener transition.\cite{Novotny2010,Bouwmeester1995}
In this paper, instead of linearly sweeping $\Delta k$, we consider a detuning that varies harmonically in time. Before doing this, we introduce the slowly varying envelope approximation to establish the formal correspondence between the mechanical oscillator system and a quantum mechanical two-level system.\\[-1ex]

\subsection{The slowly varying envelope approximation}\label{sec:SVEA}
We are interested in the dynamics of the coupled oscillators when they are tuned close to resonance ($\Delta k =0$). Therefore, we transform the equations of motion Eq.~\eqref{eq:coupledoscMatrixForm} to the basis $x_+, x_-$ and obtain
\begin{equation}
\label{eq:coupledoscMatrixFormEigen}
\left[\dd{}{t} + \gamma \d{}{t} + \Omega_0^2\right]\!
\begin{bmatrix} x_+ \\ x_- \end{bmatrix} + \begin{bmatrix} -\Omega_c^2 & - \Omega_d^2\\ -\Omega_d^2 & \;\,\Omega_c^2\end{bmatrix}\begin{bmatrix} x_+ \\ x_- \end{bmatrix}
\;=\; \begin{bmatrix} f(t) \\ f(t) \end{bmatrix}\, ,
\end{equation}
where we have used the transformation matrix $U(\Delta k=0)$ given in Eq.~\eqref{eq:TrafoUres}.
Note that the transformation to the (resonant) eigenmodes has interchanged the roles of detuning and coupling in the matrix in Eq.~\eqref{eq:coupledoscMatrixFormEigen} as compared to Eq.~\eqref{eq:coupledoscMatrixForm}. A driving force $f(t)$ can be used to excite the system in any eigenmode or superposition of eigenmodes. However, since we are interested in the dynamics of the system after its initialization, we will from now on set $f(t)=0$.\\[-1ex]

To understand the evolution of the eigenmodes we write
\begin{equation}
\label{eq:SVEAansatz}
\begin{aligned}
x_+ & \,=\,\text{Re}\left\{a(t)\exp{[\imu\Omega_0t]}\right\}\\
x_- & \,=\,\text{Re}\left\{b(t)\exp{[\imu\Omega_0t]}\right\},
\end{aligned}
\end{equation}
where each mode is rapidly oscillating at the carrier frequency $\Omega_0$ and modulated by the slowly varying complex amplitudes $a(t)$ and $b(t)$, respectively. Upon inserting Eq.~\eqref{eq:SVEAansatz} into the coupled equations of motion~\eqref{eq:coupledoscMatrixFormEigen} we assume that the amplitude functions $a(t)$ and $b(t)$ do not change appreciably during an oscillation period $2\pi/\Omega_0$, which allows us to neglect terms containing second time derivatives. This approximation corresponds to the {\em slowly varying envelope approximation} (SVEA). Furthermore, since we consider weak damping we use  $2\imu\Omega_0+\gamma\approx 2\imu\Omega_0$. Using these approximations we arrive at the following equations of motion for the eigenmode amplitudes
\begin{equation}\label{eq:SVEAsystem}
\imu\begin{bmatrix} \dot{a} \\ \dot{b}\end{bmatrix} \;=\; \frac{1}{2}\begin{bmatrix} \Delta\Omega-\imu\gamma & \omega_d\\  \omega_d & -\Delta\Omega-\imu\gamma \end{bmatrix} \begin{bmatrix} a\\b\end{bmatrix}.
\end{equation}
To simplify notation we have introduced the rescaled detuning frequency
\begin{equation}
\label{eq:rescaledDetuning}
\omega_d=\Omega_d^2/\Omega_0.
\end{equation}
Note that, for the case of vanishing damping ($\gamma=0$), Eq.~\eqref{eq:SVEAsystem} resembles the time dependent Schr\"{o}dinger equation $\imu\hbar\:\!\partial_t\ket{\Psi}=\hat{H}\ket{\Psi}$ for a state vector $\ket{\Psi}=a(t)\ket{g}+b(t)\ket{e}$ that is a superposition of a ground state $\ket{g}$ and an excited state $\ket{e}$ separated in energy by $\hbar\Delta\Omega$.  The two states are coupled by $\bra{e}\hat{H}\ket{g}=\hbar\omega_d/2$. Accordingly, our system of coupled harmonic oscillators can be considered as a ``mechanical atom'' whose ground state (excited state) is represented by the symmetric eigenmode $x_+$ (antisymmetric eigenmode $x_-$). Importantly,  the detuning $\Delta k$ of the oscillators leads to a coupling of the eigenmodes $x_+$ and $x_-$.

\subsection{Parametrically driven system}\label{sec:paramDetunedSystem}
We now investigate the dynamics of the ``mechanical atom" for a time harmonic detuning
\begin{equation}
\label{eq:parametricDriving}
\Delta k(t)\;=\;-2\Omega_0m\!\:A\:\!\cos[\omega_\text{drive}t]\; ,
\end{equation}
such that $\omega_d = -A \left(\exp[\imu\omega_\text{drive}t]+\exp[-\imu\omega_\text{drive}t] \right)$. The amplitude $A$ corresponds to the magnitude of the external modulation, whereas $\omega_{\rm drive}$ is the frequency of the modulation (c.f. Fig.~\ref{fig:coupledoscillators}).
To ease the notation we apply the transformation
\begin{equation}
\begin{aligned}
\label{eq:RotFrameAnsatz}
a&\;=\;\bar{a}(t)\exp{[-\imu\frac{\omega_\text{drive}}{2}t]}\\
b&\;=\;\bar{b}(t)\exp{[+\imu\frac{\omega_\text{drive}}{2}t]}.
\end{aligned}
\end{equation}
Here, $\bar{a}$ and $\bar{b}$ are the slowly varying amplitudes of the symmetric and antisymmetric eigenmodes  in a coordinate frame rotating at the driving frequency.
This transformation generates terms $\exp[\pm3\:\!\imu\:\!\omega_\text{drive}t/2]$ in Eq.~\eqref{eq:SVEAsystem} which are rapidly oscillating and which we neglect since they average out on the time scales of interest. This approximation is commonly referred to as the {\em rotating wave approximation} (RWA).
In the RWA and after transformation into the rotating coordinate frame Eq.~\eqref{eq:SVEAsystem} reads
%\begin{equation}
%\label{eq:SVEAsystemRWA}
%\imu\begin{bmatrix} \dot{a} \\ \dot{b}\end{bmatrix} = \frac{1}{2}\begin{bmatrix} \Delta\Omega-\imu\gamma & \frac{1}{2}\frac{A}{\Omega_0}\exp[-\imu\omega t]\\  \frac{1}{2}\frac{A}{\Omega_0}\exp[-\imu\omega t] & -\Delta\Omega-\imu\gamma \end{bmatrix} \begin{bmatrix} a\\b\end{bmatrix},
%\end{equation}
%
\begin{equation}
\label{eq:SVEAsystemRotFrame}
\imu\begin{bmatrix} \dot{\bar{a}} \\ \dot{\bar{b}}\end{bmatrix} \;=\; \frac{1}{2}\begin{bmatrix} \delta-\imu\gamma \,& \;-A\\  -A \,& -\delta-\imu\gamma \end{bmatrix} \begin{bmatrix} \bar{a}\\\bar{b}\end{bmatrix},
\end{equation}
where we have defined the detuning $\delta$ between the driving frequency and the level splitting
\begin{equation}
\label{eq:detuning}
\delta=\Delta\Omega-\omega_\text{drive}.
\end{equation}
Note that the RWA and the transformation into a rotating coordinate frame have turned the problem of two parametrically driven modes into a simple problem of two modes with a static coupling.
With the initial conditions $\bar{a}(t=0)=\bar{a}_0$ and $\bar{b}(t=0)=\bar{b}_0$ the solutions of Eq.~\eqref{eq:SVEAsystemRotFrame} are
\begin{equation}
\label{eq:solution_ab}
\begin{aligned}
\bar{a}(t)&=\left[\imu\frac{A}{\Omega_R}\sin\left(\frac{\Omega_Rt}{2}\right)\,\bar{b}_0\right. \\&+ \left. \left\{\cos\left(\frac{\Omega_Rt}{2}\right)-\imu\frac{\delta}{\Omega_R}\sin\left(\frac{\Omega_Rt}{2}\right)\right\}\bar{a}_0\right]\exp\left[-\frac{\gamma}{2}t\right],\\
\bar{b}(t)&=\left[\imu\frac{A}{\Omega_R}\sin\left(\frac{\Omega_Rt}{2}\right)\,\bar{a}_0\right. \\&+ \left. \left\{\cos\left(\frac{\Omega_Rt}{2}\right)+\imu\frac{\delta}{\Omega_R}\sin\left(\frac{\Omega_Rt}{2}\right)\right\}\bar{b}_0\right]\exp\left[-\frac{\gamma}{2}t\right],
\end{aligned}
\end{equation}
where we have introduced the generalized Rabi-frequency
\begin{equation}\label{eq:RabiFreq}
\Omega_R=\sqrt{A^2+\delta^2}.
\end{equation}
Equations~\eqref{eq:solution_ab} together with~\eqref{eq:Res_eigenmodes}, \eqref{eq:SVEAansatz} and \eqref{eq:RotFrameAnsatz} are the general solutions to the problem of two coupled harmonic oscillators under a time harmonic detuning.

\subsection{The Bloch sphere and the Bloch equations}
\begin{figure}[!b]
\includegraphics[width=\figwidth]{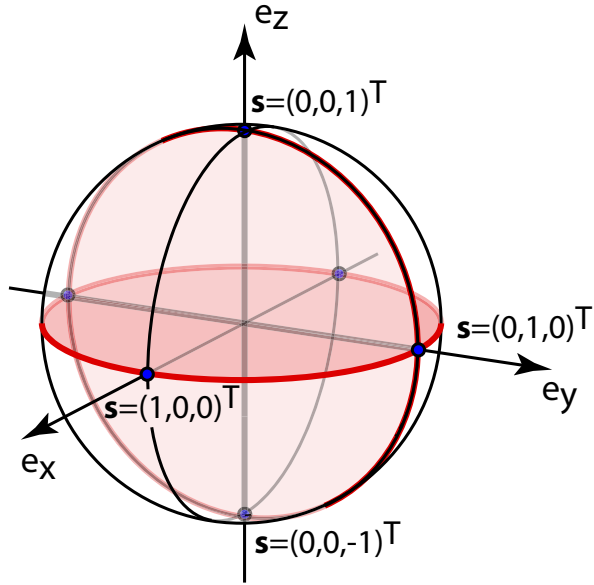}
%\epsfysize=19em
%\centerline{\epsfbox{./Fig_BlochSphere.eps}}
\caption{The Bloch sphere. A pair of amplitudes ($\bar{a},\bar{b}$) is represented by the Bloch vector $\vect{s}=(s_x,s_y,s_z)^T$.  Normalized amplitude pairs  ($|\bar{a}|^2+|\bar{b}|^2=1$) lie on the surface of the Bloch sphere, which has unit radius. All amplitude pairs $(\bar{a},\bar{b})\exp[\imu\varphi]$ with arbitrary $\varphi$ are mapped onto the same point $\vect{s}$.}
\label{fig:BlochSphere}
\end{figure}
While we have found the general solutions to the equations of motion for the parametrically driven system, it is worth representing these solutions in a vectorial form that was originally introduced by Felix Bloch in the context of nuclear magnetic resonance (NMR).\cite{Bloch1946}
We introduce the Bloch vector $\vect{s}=(s_x,s_y,s_z)^T$ with the components
\begin{equation}
\label{eq:blochvectoreq}
\begin{aligned}
s_x &=& \bar{a} \,\bar{b}^{\ast}  +  \bar{a}^{\ast} \bar{b} &=& 2\, {\rm Re}\{\bar{a} \,\bar{b}^{\ast}\} &=& 2|\bar{a}||\bar{b}|\cos(\phi)\\
s_y &=& \imu\,  \bar{a} \,\bar{b}^{\ast} - \imu \, \bar{a}^{\ast} \bar{b} &=& -2\;\! {\rm Im}\{\bar{a} \,\bar{b}^{\ast}\} &=& -2|\bar{a}||\bar{b}|\sin(\phi)\\
s_z &=&  \bar{a}\, \bar{a}^{\ast} -  \bar{b}\, \bar{b}^{\ast} &=& |\bar{a}|^2 - |\bar{b}|^2\; .
\end{aligned}
\end{equation}
The Bloch vector $\vect{s}$ encodes in its three real-valued components the state of the coupled oscillators, which is represented by the amplitudes $|\bar{a}|, |\bar{b}|$ and the relative phase $\phi$. Importantly, every state $(\bar{a},\bar{b})$ of the oscillator system can be multiplied by an arbitrary phase factor $\exp[\imu\varphi]$ without changing the corresponding Bloch vector $\vect{s}$. Discarding this absolute phase of the complex amplitudes $\bar{a},\bar{b}$ reduces the degrees of freedom from four (two real amplitudes and two phases for $\bar{a}$ and $\bar{b}$) to three, such that the state of the oscillator system can be represented in the three dimensional Bloch vector space.\\[-1ex]

For an undamped system ($\gamma=0$) and appropriately normalized amplitudes ($|\bar{a}|^2+|\bar{b}|^2=1$) the tip of the Bloch vector always lies on a unit sphere, called the Bloch sphere. For the sake of example, let us consider a few distinct points on the Bloch sphere, sketched in Fig.~\ref{fig:BlochSphere}. The north pole of the Bloch sphere $\vect{s}=(0,0,1)^T$ corresponds to the state vector $(\bar{a},\bar{b})=(1,0)$. In this state only the symmetric eigenmode of the system is excited, that is, the mechanical atom is in its ground state. Accordingly, when only the antisymmetric eigenmode is excited and the mechanical atom is in its excited state, corresponding to $(\bar{a},\bar{b})=(0,1)$, the tip of the Bloch vector is located at the south pole of the Bloch sphere $\vect{s}=(0,0,-1)^T$. All points on the equator of the Bloch sphere correspond to equal superpositions of the two eigenmodes, but with with varying relative phase $\phi$. For example, the state $(\bar{a},\bar{b})=(1,1)/\sqrt{2}$ lies at the intersection of the $\vect{e}_x$-axis and the Bloch sphere $\vect{s}=(1,0,0)^T$, whereas the state $(\bar{a},\bar{b})=(1,\imu)/\sqrt{2}$ lies at the intersection with the $\vect{e}_y$-axis $\vect{s}=(0,1,0)^T$.\\[-1ex]

It is instructive to express the dynamics of the coupled oscillator system in terms of the Bloch vector $\vect{s}$.
%We are now interested in the time evolution of the mechanical atom expressed in terms of its Bloch vector $\vect{s}$.
Using Eqs.~\eqref{eq:SVEAsystemRotFrame} and~\eqref{eq:blochvectoreq} we can easily show that the time evolution of the Bloch vector is given by
\begin{equation}\label{eq:BlochEqMotion}
\frac{\diff}{\diff t}\begin{bmatrix} s_x \\ s_y \\ s_z\end{bmatrix} = \begin{bmatrix} -\gamma & -\delta & 0 \\  \delta & -\gamma & A \\ 0 & -A & -\gamma \end{bmatrix} \begin{bmatrix} s_x \\ s_y \\ s_z\end{bmatrix} .
\end{equation}
This system of equations can be represented in compact form as
\begin{equation}
\label{eq:BlochEqMotion2}
\dot{s} \;=\; \vect{R} \times \vect{s} - \gamma\vect{s},
\end{equation}
where we defined the rotation vector $\vect{R}=(-A, 0, \delta)^T$. The equation of motion $\dot{\vect{s}}=\vect{R}\times\vect{s}$ describes the precession of the Bloch vector $\vect{s}$ around the rotation vector $\vect{R}$ with the angular frequency  $\Omega_R$ defined in Eq.~\eqref{eq:RabiFreq}.  $\Omega_R$  equals the length of $\vect{R}$.\cite{Allen1987} Equations~\eqref{eq:BlochEqMotion} define the \emph{classical Bloch equations} and are the main result of this paper. The classical Bloch equations are formally identical to the quantum Bloch equations~\cite{Allen1987} with the exception of the damping terms.\\[-1ex]

Let us briefly recap here. We are dealing with a system of two coupled harmonic oscillators. The state of this system is entirely defined by the complex amplitudes of its eigenmodes, which correspond to a distinct points on the Bloch sphere. According to Eq.~\eqref{eq:BlochEqMotion2} we can bring the system from any starting point to any other point on the Bloch sphere simply by choosing the right rotation vector $\vect{R}$ and waiting for the right time to achieve the desired amount of rotation. This idea is at the core of the concept of \emph{coherent control}. Importantly, the rotation vector $\vect{R}$ enabling such coherent control is entirely defined by the parametric driving applied to the spring constants of the system. Remember that $A$ is nothing else but the (rescaled) amplitude of the spring modulation $\Delta k$, while $\delta$ is the difference between the frequency at which we modulate the spring constant and the frequency splitting of the mechanical atom $\Delta\Omega$. In Fig.~\ref{fig:splitting}b we have visualized the coherent redistribution of amplitudes between the eigenmodes by a parametric driving of $\Delta k$.\\[-1ex]

%It is furthermore very instructive to remind ourselves of the correspondences between our mechanical system and a quantum mechanical two-level system such as an atom. We can associate the symmetric eigenmode of the coupled oscillator system with the ground state of a two-level system and the antisymmetric eigenmode with the excited state. In this picture the transition frequency of the two-level system corresponds to the frequency splitting between the eigenmodes of the mechanical system, which is proportional to the coupling rate between the individual oscillators. The complex amplitudes $a,b$ of the eigenmodes of the mechanical system correspond to the complex amplitudes of the wavefunction written as a superposition of ground and excited state of the two-level system. Accordingly, the populations $|a|^2,|b|^2$ correspond to the populations of the quantum mechanical states.

Having pointed out the similarities between a mechanical oscillator system and a quantum mechanical two-level system we note an important difference. The damping $\gamma$ ultimately forces the system into the state $(\bar{a},\bar{b})=(0,0)$, which means that the Bloch vector disappears. In contrast, due to spontaneous emission a quantum two-level system will always end up in its ground state after a long time. Clearly, while the concept of coherent control can be applied to entirely classical systems, spontaneous emission is a process that is genuinely quantum mechanical in nature and cannot be recovered in a system governed by classical mechanics.
It must be emphasized that spontaneous emission requires a fully quantized theory and cannot be derived by semi-classical quantum mechanics. Even Bloch added the decay constants semi-phenomenologically in his treatment of nuclear spins.\cite{Allen1987} In quantum mechanical systems one commonly distinguishes two decay rates. The first one, termed longitudinal decay rate $\gamma_1=1/T_1$, describes the decay of the inversion $s_z$, while the second one, the transverse decay rate $\gamma_2=1/T_2$ governs the decay of the components $s_x$ and $s_y$ of the Bloch vector. For the mechanical oscillator system we find $\gamma_1=\gamma_2=\gamma$ which explains the recent experimental finding by Faust \emph{et al.} in a micromechanical oscillator system.\cite{Faust2013}
Finally, we stress that the analogy between a pair of coupled mechanical oscillators and a quantum two-level system relies on the SVEA and accordingly breaks down whenever the amplitudes $|\bar{a}|^2, |\bar{b}|^2$ change on a time scale comparable to the carrier frequency $\Omega_0$.\\[-1ex]

\section{Operations on the Bloch sphere}
We have seen that by parametrically modulating the detuning of the coupled oscillators we can control the trajectory of the Bloch vector on the Bloch sphere at will. Three protocols for Bloch vector manipulation have proven particularly useful in the field of coherent control, leading to phenomena called Rabi oscillations, Ramsey fringes, and Hahn echo. We will now briefly discuss these phenomena.

\subsection{Rabi oscillations}
In 1937 Rabi studied the dynamics of a magnetic spin in a static magnetic field that is modulated by a radio frequency field and he found that the spin vector is periodically oscillating between parallel and anti-parallel directions with respect to the static magnetic field.\cite{Rabi1937} These oscillations are referred to as Rabi oscillations, or Rabi flopping. The mechanical atom reproduces the basic physics of this Rabi flopping.\\[-1ex]

%, with the exception of damping associated with spontaneous emission.
\begin{figure}[!b]
\includegraphics[width=\figwidth]{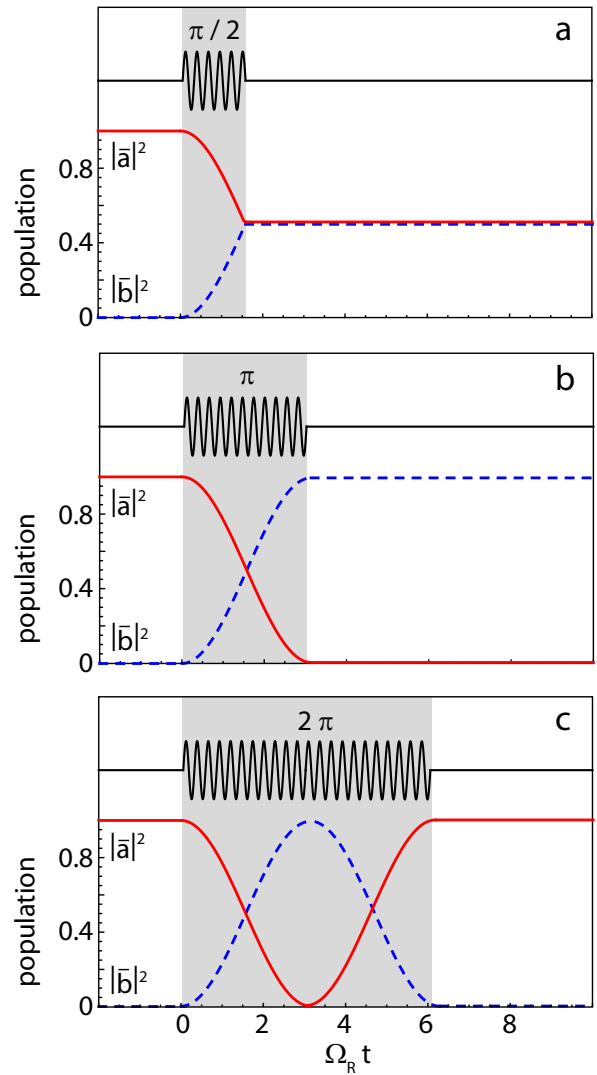}
%\vspace{1.5em}
%\epsfysize=38em
%\centerline{\epsfbox{./Fig_Pulses.eps}\vspace{-1.5em}}
\caption{Controlling the dynamics of  coupled oscillators with  pulses of different duration and amplitude. (a) Starting in the ground state $(\bar{a},\bar{b})=(1,0)$ a $\pi/2$-pulse leaves both modes equally excited. (b) A $\pi$-pulse transfers the energy of $x_+$ to $x_-$, and (c) a $2\pi$ pulse brings the system back to where it started. In all cases we assumed no damping ($\gamma=0$) and no detuning ($\delta=0$).
\label{fig:pulses}}
\end{figure}
\begin{figure*}[htb]
\includegraphics[width=1.4\columnwidth]{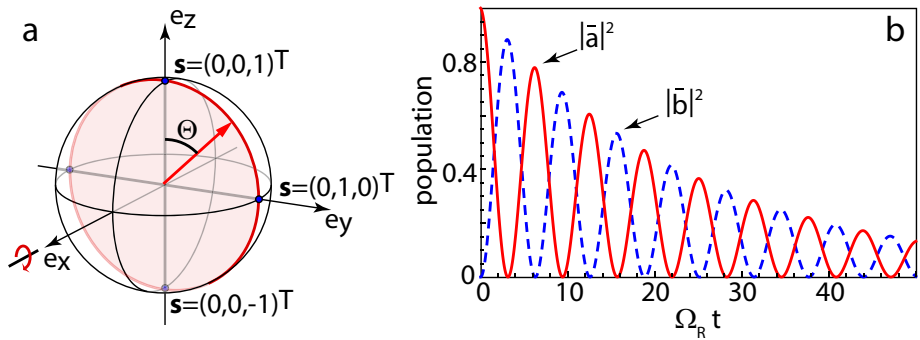}
%\epsfysize=15em
%\centerline{\epsfbox{./Fig_RabiOsc.eps}\vspace{-1.5em}}
\caption{(a) Bloch sphere with trajectory of Bloch vector during resonant Rabi oscillations marked in red. Starting from the north pole the Bloch vector rotates around the $\vect{e}_x$-axis.  To rotate the Bloch vector by the angle $\Theta$ the driving field with amplitude $A$ has to be turned on for a time $t_\Theta=\Theta/A$. (b) Rabi oscillations of the populations $|\bar{a}|^2,|\bar{b}|^2$ for zero detuning ($\omega_\text{drive}=\Delta\Omega$) and damping $\gamma=\Omega_R/25$. The energy flops back and forth between the two oscillation modes $x_+$ and $x_-$. The Rabi frequency $\Omega_R$ defines the flopping rate, and is given by the rescaled modulation amplitude $A$ of the detuning $\Delta k$.
\label{fig:rabiosc}}
\end{figure*}
Let us neglect damping for the moment ($\gamma=0$) and assume a resonant ($\delta=0$) parametric driving $\Delta k \propto A\cos(\Delta\Omega\, t)$ to our system, such that the Bloch vector, starting at the north pole $\vect{s}=(0,0,1)^T$, rotates around the axis $\vect{R}=-A\:\!\vect{e}_x$ at a frequency $\Omega_R=A\,$ according to Eq.~\eqref{eq:BlochEqMotion2}. After a time $t_\pi=\pi/A$ the Bloch vector will have rotated to the south pole $\vect{s}=(0,0,-1)^T$. This means that the symmetric eigenmode now has zero amplitude while the antisymmetric eigenmode has unit amplitude, that is, $(\bar{a},\bar{b})=(0,1)$. Obviously, parametric driving for a time $t_\pi$ (called $\pi$-pulse) inverts our system.  Accordingly, after parametrically driving the system for a time $t_{2\pi}=2\pi/A$ it has returned to its initial state at the north pole of the Bloch sphere. We have plotted the populations of the eigenmodes for three different pulse durations in Fig.~\ref{fig:pulses}.
In general, for a $\Theta$-pulse the parametric driving with amplitude $A$ is turned on for a time $t_\Theta=\Theta/A$. Remember that the driving signal oscillates at a frequency $\omega_\text{drive}$ according to Eq.~\eqref{eq:parametricDriving} and therefore undergoes many oscillations during the pulse. Importantly, we can make every pulse arbitrarily short by applying a driving signal with large amplitude $A$.
For a continuous parametric driving, starting at $\vect{s}=(0,0,1)^T$, the system is oscillating between its two eigenmodes at the resonant Rabi-frequency $\Omega_R=A$. We can explicitly convince ourselves that the picture of the Bloch sphere yields the correct result by considering the time evolution of the population of the eigenmodes, given by $|\bar{a}|^2$ and $|\bar{b}|^2$ in Eq.~\eqref{eq:solution_ab}. For $(\bar{a}_0,\bar{b}_0)=(1,0)$ we obtain
\begin{equation}
\label{eq:solution_Rabi}
\begin{aligned}
|\bar{a}(t)|^2&\;=\;\cos^2\!\left(\Omega_Rt/2\right)\,\exp\left[-\gamma t\right]\\
|\bar{b}(t)|^2&\;=\;\sin^2\!\left(\Omega_Rt/2\right)\,\exp\left[-\gamma t\right].
\end{aligned}
\end{equation}
We plot the trajectory of the Bloch vector for a resonantly driven system in Fig.~\ref{fig:rabiosc}a and the populations $|\bar{a}|^2,|\bar{b}|^2$ in Fig.~\ref{fig:rabiosc}b. Indeed, for zero damping ($\gamma=0$) the energy  oscillates back and forth between the two eigenmodes of the system at a frequency $\Omega_R$. However, a finite damping $\gamma$ makes the population of both eigenmodes die out with progressing time and the length of the Bloch vector is no longer conserved. For finite detuning ($\delta\ne0$) we find that even without damping the population inversion is reduced and no longer reaches a value of one. The population of the antisymmetric eigenmode reaches a maximum value of $|\bar{b}(t_\pi)|^2=A^2/\Omega_R^2$. The fact that the Rabi oscillations do not lead to a complete inversion of the system for finite detuning can be understood by considering the rotation of the Bloch vector on the Bloch sphere. For finite detuning the rotation vector $\vect{R}$ has a component in $\vect{e}_z$-direction such that a rotation of the Bloch vector $\vect{s}$ starting at the north pole of the Bloch sphere no longer reaches the south pole. Another important observation is that the non-resonant Rabi-oscillations always proceed at a frequency larger than in the resonant case ($\Omega_R>\Omega_R^\text{res}$ for all $\delta\ne0$) according to Eq.~\eqref{eq:RabiFreq}.\\[-1ex]

\subsection{Ramsey fringes}
We have seen that Rabi oscillations correspond to rotations around the $\vect{e}_x$-axis. What about rotations around other axes? Clearly, any detuning $\delta$ leads to an $\vect{e}_z$-component of the rotation vector $\vect{R}$, that is, a rotation around the $\vect{e}_z$-axis. Assume we prepare the mechanical oscillator system in a state $\vect{s}=(0,1,0)^T$. As we know from the previous section, we can reach this point by applying a $\pi/2$-pulse to a system in the state $\vect{s}=(0,0,1)^T$.
If we allow for a finite detuning $\delta$, according to Eq.~\eqref{eq:BlochEqMotion} the system will evolve away from the state $\vect{s}=(0,1,0)^T$ even if the driving is off ($A=0$). In fact, for $A=0$ the Bloch vector rotates around the Bloch sphere's equator at an angular frequency $\delta$, progressing by an angle $\Phi=\delta t$ in the equatorial plane within a time $t$. \\[-1ex]

It is puzzling at first sight that the Bloch vector is rotating even though no driving is applied to the system. To understand the precession of the Bloch vector for a finite detuning it is necessary to remind ourselves of two important facts: the first one is the frequency difference $\Delta\Omega$ between the eigenmodes of the system, the second one are the coordinate transformations we applied.
\begin{figure*}[htb]
\includegraphics[width=1.4\columnwidth]{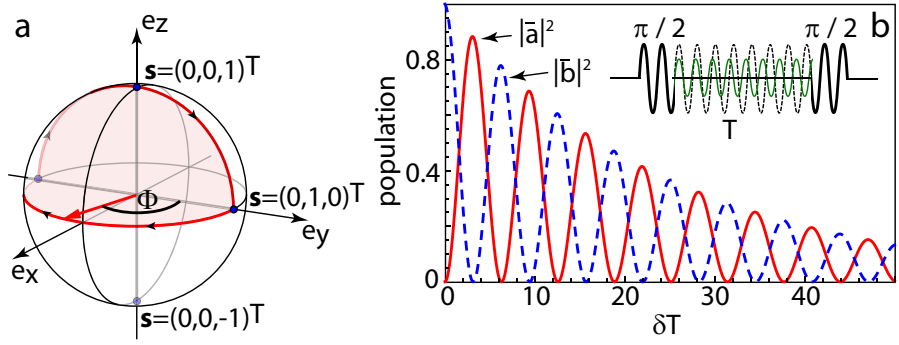}
%\epsfysize=17em
%\centerline{\epsfbox{./Fig_RamseyOsc.eps}\vspace{-0.0em}}
\caption{(a) Trajectory of Bloch vector in a Ramsey experiment with detuning $\delta$ and waiting time $T=\pi/\delta$. During this time the Bloch vector makes a half rotation on the equator. During the time $t=\Phi/\delta$ the Bloch vector precesses by an angle $\Phi$ in the equatorial plane. (b) Ramsey fringes for finite damping $\gamma\ne0$. The populations $|\bar{a}|^2,|\bar{b}|^2$ are inverted after $T=n\,\pi/\delta$,  with $n$ being an integer. The inset illustrates the applied pulse sequence which consists of two $\pi/2$-pulses separated by a waiting time $T$. The Ramsey fringes result from a phase difference acquired by the (mechanical) atom, which is due to the drive frequency $\omega_\text{drive}$ (dashed in inset while driving is off) being different from the transition frequency $\Delta\Omega$ (green).
\label{fig:ramseyosc}}
\end{figure*}
In the transformation Eq.~\eqref{eq:SVEAansatz} we separated out all fast oscillations at the carrier frequency $\Omega_0$. Accordingly, since the eigenmodes $x_+, x_-$ oscillate at frequencies $\Omega_\pm$, the amplitudes $a, b$ still contain oscillations at $\pm\Delta\Omega/2$. This frequency difference between the eigenmodes means that as time passes they acquire a relative phase difference. By transforming into a coordinate system rotating at the driving frequency in Eq.~\eqref{eq:RotFrameAnsatz} we exactly compensate for the phase difference between the eigenmodes if the driving frequency $\omega_\text{drive}$ exactly corresponds to the frequency splitting $\Delta\Omega$ between the levels. Thus, in this rotating frame, the Bloch vector is constant. If, however, we choose a finite $\delta$, we are transforming into a reference system that is detuned with respect to the transition frequency $\Delta\Omega$ and the Bloch vector will rotate even in the absence of driving.
The detuning $\delta$ can be made visible in an experimental scheme devised by Ramsey.\cite{Ramsey1950} The Ramsey method consists of three steps, as illustrated in the inset of Fig.~\ref{fig:ramseyosc}b. First we bring the system from the north pole of the Bloch sphere to the state $\vect{s}=(0,1,0)^T$ with a short $\pi/2$-pulse with detuning $\delta$, second we wait for a time $T$, and finally we apply another $\pi/2$-pulse. In case of zero detuning the Bloch vector has not moved during the waiting time and accordingly we end up on the south pole of the Bloch sphere. However, if the detuning is finite, the Bloch vector will precess around the $\vect{e}_z$-axis by an angle $\Phi=\delta T$ before being rotated by $\pi/2$ around the $\vect{e}_x$-axis by the second pulse. For example, after a time $T=\pi/\delta$ the Bloch vector will have rotated to the point $\vect{s}=(0,-1,0)^T$ such that the second $\pi/2$-pulse will bring the system back to the north pole $\vect{s}=(0,0,1)^T$, as plotted in Fig.~\ref{fig:ramseyosc}a. If we measure the populations $|a|^2,|b|^2$ after the second $\pi/2$-pulse we find the characteristic Ramsey fringes, plotted in Fig.~\ref{fig:ramseyosc}b, where we also included a finite damping rate $\gamma$ that leads to an exponential decay of both populations.

\commentOut{
We can intuitively understand why the second $\pi/2$-pulse brings the system down to the south pole $(0,1)$ for the resonant case but back up to $(1,0)$ for a detuning $\delta=\pi/T$. The reason is that in the resonant case the driving field (despite being off during the time $T$) has not acquired a phase difference with respect to the mechanical atom oscillating at its transition frequency $\Delta\Omega$. Accordingly, whatever the waiting time $T$, the next pulse will simply continue the movement on the Bloch sphere. In contrast, in the case of finite detuning, the driving (illustrated as the solid black line while on in the inset of Fig.~\ref{fig:ramseyosc}b and as a dashed line when off) and the mechanical atom (whose oscillation at frequency $\Delta\Omega$ is sketched as the green line in the inset of Fig.~\ref{fig:ramseyosc}b) acquire a phase difference during the time $T$. If that phase difference equals $\pi$ the second pulse deexcites the mechanical atom back to the initial state $(1,0)$ instead of further exciting it to $(0,1)$.\\[-1ex]
}

\begin{figure*}[htb]
\includegraphics[width=1.4\columnwidth]{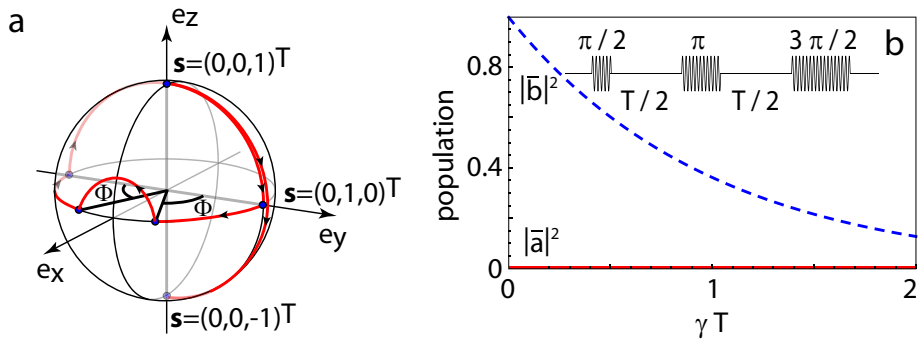}
%\epsfysize=13em
%\centerline{\epsfbox{./Fig_HahnEcho.eps}
%\vspace{-1em}}
\caption{Hahn echo experiment. (a) The sequence of pulses and wait times represented as a trajectory on the Bloch sphere. The path starts at the north pole $\vect{s}=(0,0,1)^T$ and ends at the south pole $\vect{s}=(0,0,-1)^T$. (b) Measurement outcome of the Hahn experiment plotted as the populations $|\bar{a}|^2,|\bar{b}|^2$. The population of the symmetric mode $x_+$ is zero for all delay times $T$ while the population of the antisymmetric mode $x_-$ falls off exponentially with the damping constant $\gamma$. Inset: Illustration of the pulse sequence.
\label{fig:hahnecho}}
\end{figure*}
\subsection{Hahn-echo}
Having established that a sequence of pulses and wait times can be represented by a trajectory on the Bloch sphere, we now consider the dynamics of the so-called \emph{Hahn echo} experiment.\cite{Hahn1950} The experiment involves a sequence of three short pulses (see inset of Fig.~\ref{fig:hahnecho}b).
As before, we start with the system in the lower energy eigenmode corresponding to the Bloch vector $\vect{s}=(0,0,1)^T$. Similar to the Ramsey experiment, we bring the system with a short $\pi/2$-pulse to $\vect{s}=(0,1,0)^T$, which corresponds to a superposition of the $x_+$ and $x_-$ modes.
During a waiting time $T/2$ the Bloch vector rotates around the $\vect{e}_z$-axis according to the detuning $\delta$ by an angle $\Phi=\delta T/2$ until one applies a short $\pi$-pulse. This pulse flips the Bloch vector around the $\vect{e}_x$-axis as illustrated in Fig.~\ref{fig:hahnecho}a. Another wait time of $T/2$ rotates the Bloch vector to $\vect{s}=(0,-1,0)^T$, from where a $3\pi/2$ pulse brings it to the south pole $\vect{s}=(0,0,-1)^T$.
However, this is only true without damping. The complex amplitudes $(\bar{a},\bar{b})$ of the oscillation modes are both damped by a factor $\exp[-\gamma t/2]$ according to Eq.~\eqref{eq:solution_ab} and hence
\begin{equation}
\begin{aligned}
|a(T)|^2 &= 0\\
|b(T)|^2 &= \text{e}^{-\gamma T}.
\label{eq:hahnresult}
\end{aligned}
\end{equation}
Thus, as shown in Fig.~\ref{fig:hahnecho}b, by varying the time delay $T$ and measuring the energy in oscillation mode $x_-$ we can experimentally determine the damping $\gamma$ in the system.
Importantly, in the Hahn protocol, the precession performed by the Bloch vector during the first waiting interval is exactly compensated during the second waiting time. This fact is extraordinarily useful if one does an ensemble measurement on several systems with a distribution of detunings or if one repeatedly measures the same system which suffers from fluctuations in its transition frequency $\Delta\Omega$.
In quantum mechanics one distinguishes two types of damping, energy relaxation and phase decoherence. Since the propagation of the Bloch vector in the Hahn echo experiment happens only in the xy-plane (the pulses can be made arbitrarily short by increasing their amplitude) the Hahn echo experiment is able to eliminate the contribution due to energy relaxation and only render the damping due to phase decoherence.~\cite{Allen1987}\\[-1ex]

\section{Conclusions}
We have shown that a pair of coupled mechanical oscillators with parametrically driven detuning can serve as a classical analogon to a quantum mechanical two-level system. The correspondence between the classical and the quantum system is established by using the slowly varying envelope approximation, which casts the Newtonian equations of motion of the coupled oscillators into a form resembling the Schr{\"o}dinger equation for a two-level atom. The mechanical oscillator analogon features only a single decay rate for both longitudinal and transverse damping in contrast to a quantum mechanical two-level system which can show different decay rates due to spontaneous emission and dephasing processes. Also, the classical model does not conserve the total population of the levels since both eigenmodes are equally damped. Our treatment provides an easy introduction to the field of coherent control in the context of both classical and quantum systems. An extension of our discussion of the mechanical atom lends itself to introduce students to Pauli matrices, density operators and rotation operators.

\begin{acknowledgments}
The authors thank  Lo\"{i}c Rondin for fruitful discussions and are most grateful for financial support by ETH Z{\"u}rich and  ERC-QMES (No. 338763).
\end{acknowledgments}

\bibliography{NovotnyFrimmer_ClassicalBlochEquations}

\begin{thebibliography}{10}
\newcommand{\enquote}[1]{``#1''}

\bibitem{Alzar2002}
C.~L.~G. Alzar, M.~A.~G. Martinez, and P.~Nussenzveig, \enquote{Classical
  analog of electromagnetically induced transparency,} Am. J. Phys.
  \textbf{70}, 37--41 (2002).

\bibitem{Shore2009}
B.~W. Shore, M.~V. Gromovyy, L.~P. Yatsenko, and V.~I. Romanenko,
  \enquote{Simple mechanical analogs of rapid adiabatic passage in atomic
  physics,} Am. J. Phys. \textbf{77}, 1183--1194 (2009).

\bibitem{Maris1988}
H.~J. Maris and Q.~Xiong, \enquote{Adiabatic and nonadiabatic processes in
  classical and quantum mechanics,} Am. J. Phys. \textbf{56}, 1114--1117
  (1988).

\bibitem{Novotny2010}
L.~Novotny, \enquote{Strong coupling, energy splitting, and level crossings: A
  classical perspective,} Am. J. Phys. \textbf{78}, 1199--1202 (2010).

\bibitem{Allen1987}
L.~Allen and J.~H. Eberly, \emph{Optical Resonance and Two-level Atoms}, Dover
  Books on Physics Series (Dover, 1987).

\bibitem{Dragoman2004}
D.~Dragoman and M.~Dragoman, \emph{Quantum-Classical Analogies}, The Frontiers
  Collection (Springer, 2004).

\bibitem{Tobar1991}
M.~E. Tobar and D.~G. Blair, \enquote{A generalized equivalent circuit applied
  to a tunable sapphire-loaded superconducting cavity,} IEEE Trans. Microw.
  Theory Techn. \textbf{39}, 1582--1594 (1991).

\bibitem{Spreeuw1990}
R.~J.~C. Spreeuw, N.~J. van Druten, M.~W. Beijersbergen, E.~R. Eliel, and J.~P.
  Woerdman, \enquote{Classical realization of a strongly driven two-level
  system,} Phys. Rev. Lett. \textbf{65}, 2642--2645 (1990).

\bibitem{Bouwmeester1995}
D.~Bouwmeester, N.~H. Dekker, F.~E.~v. Dorsselaer, C.~A. Schrama, P.~M. Visser,
  and J.~P. Woerdman, \enquote{Observation of {L}andau-{Z}ener dynamics in
  classical optical systems,} Phys. Rev. A \textbf{51}, 646--654 (1995).

\bibitem{Okamoto2013}
H.~Okamoto, A.~Gourgout, C.-Y. Chang, K.~Onomitsu, I.~Mahboob, E.~Y. Chang, and
  H.~Yamaguchi, \enquote{Coherent phonon manipulation in coupled mechanical
  resonators,} Nature Phys. \textbf{9}, 480--484 (2013).

\bibitem{Faust2013}
T.~Faust, J.~Rieger, M.~J. Seitner, J.~P. Kotthaus, and E.~M. Weig,
  \enquote{Coherent control of a classical nanomechanical two-level system,}
  Nature Phys. \textbf{9}, 485--488 (2013).

\bibitem{Frimmer2012}
M.~Frimmer and A.~F. Koenderink, \enquote{Superemitters in hybrid photonic
  systems: A simple lumping rule for the local density of optical states and
  its breakdown at the unitary limit,} Phys. Rev. B \textbf{86}, 235428--1--6
  (2012).

\bibitem{Bloch1946}
F.~Bloch, \enquote{Nuclear induction,} Phys. Rev. \textbf{70}, 460--474 (1946).

\bibitem{Rabi1937}
I.~I. Rabi, \enquote{Space quantization in a gyrating magnetic field,} Phys.
  Rev. \textbf{51}, 652--654 (1937).

\bibitem{Ramsey1950}
N.~F. Ramsey, \enquote{A molecular beam resonance method with separated
  oscillating fields,} Phys. Rev. \textbf{78}, 695--699 (1950).

\bibitem{Hahn1950}
E.~L. Hahn, \enquote{Spin echoes,} Phys. Rev. \textbf{80}, 580--594 (1950).

\end{thebibliography}

\end{document}